\documentclass[10pt,aps,twocolumn,pre,floatfix,superscriptaddress]{revtex4-1}
\usepackage{graphicx}
\usepackage{epstopdf}
\usepackage{amssymb,amsfonts,amsmath}
\usepackage{color}
\usepackage{float}
\usepackage{gensymb}
\usepackage{dcolumn}
\usepackage{bm}
\usepackage{wrapfig} 
\usepackage{xcolor}
\usepackage{mdframed}
\usepackage[many]{tcolorbox}
\usepackage{lmodern}
\usepackage{lipsum}
\usepackage{amsmath, nccmath}
\usepackage{multirow}
\usepackage{tabu}
\usepackage{float}
\usepackage[caption=false]{subfig}
\usepackage{capt-of}
\newcommand{\RN}[1]{\textup{\uppercase\expandafter{\romannumeral#1}}}
\usepackage{algorithm}
\usepackage[noend]{algpseudocode}
\usepackage{esvect}

\usepackage{mathtools}
\usepackage{siunitx}

\usepackage{color,soul}
\usepackage{xcolor}
\definecolor{SkyBlue}{RGB}{195,216,235}

\begin{document}


\title{Softening of DNA near Melting as Disappearance of an Emergent Property }

\author{Debjyoti Majumdar}
\email{debjyoti@iopb.res.in}
\affiliation{Institute of Physics, Bhubaneswar, Odisha, 751005, India}
\affiliation{Homi Bhabha National Institute, Training School Complex, Anushakti Nagar, Mumbai 400094, India}
\author{Somendra M. Bhattacharjee}
\email{somendra.bhattacharjee@ashoka.edu.in}
\affiliation{Department of Physics, Ashoka University, Sonepat, Haryana - 131029,
     India}
\date{\today}

\begin{abstract}
Near the melting transition the bending elastic constant $\kappa$, 
an emergent property of double-stranded DNA (dsDNA), is shown not 
to follow the rodlike scaling for small length $N$. The reduction 
in $\kappa$ with temperature is determined by the denatured bubbles 
for a continuous transition, e.g., when the two strands are Gaussian, 
but by the broken bonds near the open end in a Y-like configuration 
for a first-order transition as for strands with excluded volume 
interactions. In the latter case, a lever rule is operational implying 
a phase coexistence although dsDNA is known to be a single phase. 
\end{abstract}

\maketitle
\section{Introduction}    
  DNA stores the genetic information in its base sequence, but its
  functionality relies on its physical properties, like stiffness and
  length. The elastic energy of DNA packaged in a viral capsid helps
  in the injection process\cite{purohit,garcia}, energetically-costly
  bends of DNA provide sites for attachments of transcription factors
  and other enzymes\cite{hori,rees,topoiso}, while the melting of DNA
  is a vital step in polymerase chain reactions\cite{mullis}. The 
  topological constraint when double-stranded DNA (dsDNA) is viewed as
  a ribbon, leads to two independent elastic constants for twist and
  bend\cite{vologo,hager,dietler,abels}.  Both these elastic
  constants vanish on the melting of dsDNA\cite{brunet,vologo,olson},
  when the ribbon picture is lost, showing that the stiffness is an
  emergent property of the bound DNA\cite{kivelson}.  However, how
  this emergent behaviour goes away at melting is still unknown.  Here
  we determine the fundamental relation between the emergent bending
  elastic constant $\kappa$ and the fraction of broken base-pairs that
  drives the melting transition.  We show, by simulating long
  semiflexible DNA, that the relation is dependent on the order of the
  melting transition, and involve different physical mechanisms.
  
 For a continuous melting transition, as for Gaussian chains, 
 a renormalized semiflexible chain picture is valid where effective 
 $\kappa$ for long chains is renormalized non-trivially by the fraction 
 of broken bonds.  Melting is found to occur homogeneously along the 
 chain but a worm-like chain model is applicable only at low temperatures 
 where there are no broken pairs. In contrast, in the presence of excluded
 volume interaction, when melting is first order, the effective $\kappa$ 
 is found to be determined by a phase-coexistence type picture with the 
 reduction in rigidity coming mainly from the large fraction of broken 
 bonds near the open end of dsDNA.  It is not homogeneous melting, and 
 a phase-coexistence is at odds with the conventional mechanism of 
 bubble-induced melting transition.

\begin{figure}[t]
\includegraphics{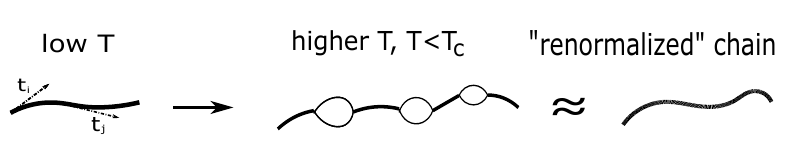}
\caption{Schematic diagram of renormalization of the elastic behaviour
  of dsDNA due to bubbles. For Gaussian chains, tangents ${\bf
    t}_i,{\bf t}_j$ can be used to define a persistence length (see
  text). Can one define a persistence length for
  configurations with bubbles?}
\label{fig:1}
\end{figure}

The stiffness of dsDNA is expressed in terms of the persistence
length\cite{hager,dietler,abels} $l_p\sim50$nm which is much larger
than that of highly flexible individual strands of DNA (ssDNA) with
$l_p\sim 2$nm. Generally, $l_p$ at temperature $T$ is defined on
dimensional ground from the bending elastic constant as
$l_p=\kappa/T$ (the Boltzmann constant $k_B=1$), whereas the intuitive
picture that a semiflexible 
polymer behaves like a rod for lengths less than $l_p$, follows from
the decay length of the tangent-tangent correlation function,
$C(i-j)=\langle{\bf t}_i\cdot{\bf t}_j\rangle\sim \exp(-|i-j|b/l_p), $
where ${\bf t}_i$ is the tangent to the space curve at monomer $i$ as
shown in Fig.~\ref{fig:1}, $b$ is the bond length, and
$\langle...\rangle$ denotes 
ensemble average.  These two definitions match for a worm-like chain
which is Gaussian at long length, but not in the presence of excluded
volume interaction when $C(i,j)$ decays as a power-law without any
typical length\cite{hager,schafer, hsu}. 

As the base-pair energy $\sim 6$-$9$ kcal/mol, thermal fluctuations
lead to a cooperative breaking of the hydrogen bonds in the long
length limit.  This is the melting of DNA\cite{vologodski3}.  The
broken base pairs may be distributed along the chain or maybe near the
open end (called the Y-fork) when one end of DNA is kept fixed. A consecutive
set of broken pairs is to be called a bubble; see Fig.~\ref{fig:2}.
This bubble mediated transition is the usual Poland-Scheraga scheme of
thermal melting of DNA\cite{fisher}.  The fraction $n_c$ of unbroken
bonds play the role of the order-parameter for the transition, viz.,
$n_c~ \neq 0\, (n_c=0)$ in the dsDNA (denatured) phase, and depending on
the nature of the interactions, the melting transition can be
continuous or first-order\cite{fisher,causo}.  As the ssDNA's are
flexible, the bubbles act as hinges for the rigid segments\cite{yan,
  forties, wiggins,vologodskii2,palsmb,bonnet,vologo}, and,
additionally, bubbles have biologically important roles
\cite{chongli,ramstein}.  The extra flexibility introduced by the
bubbles leads to a downward renormalization of the elastic constant as
shown schematically in Fig.~\ref{fig:1}, provided the bubbles are
distributed homogeneously along the chain. The loss of stiffness of
dsDNA is gradual over a range of $20\degree$C near melting
\cite{brunet}, but the validity of the homogeneous picture and the
functional form of the temperature dependence are not known. There is also a 
problem in defining $C(i,j)$ or a ``ribbon" at higher 
temperatures for configurations dotted with bubbles (Fig.~\ref{fig:1}), unless  one 
coarse-grains at the scale of the bubble size.

Why is dsDNA stiff when individual strands are not and how does that
stiffness go away with the increase of temperature?  These questions
may resemble similar ones about the rigidity of crystals.  However,
there are fundamental differences between the two cases.  For a solid,
rigidity is also an emergent phenomenon, where the shear modulus,
imparting rigidity, is a consequence of continuous-symmetry
breaking\cite{kivelson}.  There is no such scenario for $\kappa$,
especially because it is not the response function associated with any
order-parameter, like $n_c$.  Instead, the bound phase allows a
ribbon-like description for which topological arguments\cite{vologo},
e.g., the C{\u{a}}lug{\u{a}}reanu theorem, are applicable.  The twist
elastic constant, related to the helical nature, and the bending
elastic constant related to the entropy of DNA\cite{vologodski3}, are
relevant for dsDNA, but not for ssDNA or the denatured phase where the
ribbon picture is lost. Of the two elastic constants, 
$\kappa$ is a large scale property that should be insensitive to 
microscopic details, while the twist constant is dependent on the 
details of the structure.  It is, therefore, possible to model the 
reduction of $\kappa$ through the changes in the semiflexible bound 
structure mediated by the broken base-pairs.  The
occurrences of bubbles, as in Fig.~\ref{fig:1}, may seem to
invalidate the ribbon picture, even raising questions on defining a
tangent vector ${\bf t}$.  These issues may be alleviated by
coarse-graining on a scale larger than the bubble size (Fig.~\ref{fig:1}) 
restoring the ribbon picture with renormalized elastic
constants.  With this in mind, we use coarse-grained models for finite 
length DNA, which are important from experimental point of view, since 
experiments are performed upon finite system.

This paper is organized in the following manner. In 
Sec.~${\RN 2}$ the coarse-grained models are defined. Two models, viz., Gaussian 
chain model and chains with self-  and mutual-avoidance, both on a cubic 
lattice are defined there.  The connection of the elastic constant with 
appropriate sizes via fluctuation theorems are also elaborated there. Sec.~${\RN 3}$ 
gives the details of the simulation method of developing \texttt{PERM} for 
dsDNA. Sec.~${\RN 4}$ discusses the drastic difference in the behaviour of rigidity 
when excluded volume interactions are taken into account. Sec.~${\RN 5}$ concludes 
the paper with some analogies between DNA melting and crystal melting.

\section{Model and Qualitative Description}
\begin{figure}[t]
\includegraphics{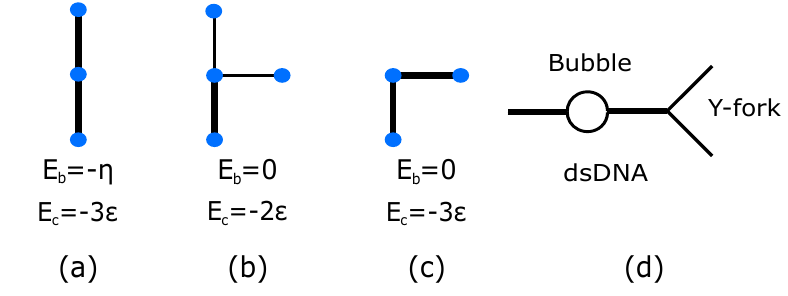}
\caption{ (a-c) Possible configurations for two successive 
bonds. (a) Two bound bonds with three contacts, and angle $\theta=0$. (b) 
Opening of a fork. (c) same as (a) but with a bend ($\theta=\pi/2$), costing 
energy. There are reverse steps for Model I. $E_c$ and $E_b$ represent total 
contact and bending energies. (d)  Identifying bubbles and the Y-fork.}
\label{fig:2}
\end{figure}

Two different models are considered here, viz., $\RN{1}$ and $\RN{2}$,
on a cubic lattice.  In model $\RN{1}$, the strands are Gaussian
chains, while in model $\RN{2}$, we incorporate both self and mutual
avoidance.  The binding of the chains is allowed by an attractive
energy $-\epsilon (\epsilon>0)$ whenever a monomer of one chain is in
contact with a monomer of the other chain provided both monomers have
the same position along the chain.  This ensures the native
base-pairing of DNA. In each case, we consider two varieties of
polymers, viz., (i) flexible polymers where both the single and double
strands are flexible, and (ii) semiflexible dsDNA where only the bound
parts are semiflexible but the bubbles consist of flexible chains.
The semi-flexibility in dsDNA is incorporated by penalizing a bend
of two successive paired bonds with energy $E_{\rm b}=-\eta\cos{\theta}$
where $\theta$ is the angle between two bonds and $\eta (\textgreater 0)$
is the bending energy constant; see Fig.~\ref{fig:2}(a)-(c). Whereas a bent
ds configuration as in Fig.~\ref{fig:2}(c) is energetically favourable
compared to Fig.~\ref{fig:2}(b), the latter is a source of additional
entropy.  Consequently, bubbles (Fig.~\ref{fig:2}(d)) are to be
expected at higher temperatures vis-a-vis bent ds-chains at lower
temperatures \cite{comm3}.

To explore the elastic behaviour, a force ${\bf F}$ is applied at the
end point ${\bf r}_i(N)$ of each strand $i=1,2$ of length $N$, keeping
the other ends fixed. The additional force-term in the Hamiltonian is
$H_F=-\bf{F}\cdot\bf{x}$, where ${\bf x}={\bf r}_1(N)+{\bf r}_2(N)$.
The elastic response can be defined from a tensorial quantity $\chi$
as $\chi_{ij}=\frac{\partial \langle x_i\rangle}{\partial F_j}$, with
the subscripts $i,j$ denoting the Cartesian components.  In the zero
force limit (${\bf F}\to 0$), isotropy can be used to define the
elastic constant as $\kappa={\rm Tr}\left[\chi \right]$, which can be
related to the zero-force fluctuations of ${\bf x}$ as $
\bar\kappa\equiv k_BT \kappa=
(\langle\textbf{x}^{2}\rangle-\langle\textbf{x}\rangle^{2}), $ where
the averaging is done with the ${\bf F}=0$ Hamiltonian.  This
fluctuation relation allows us to determine $\kappa$ without any
external force. As we see, the bending elastic constant is 
not the response function associated with $n_c$ and so the conventional
critical behaviour of response functions in phase transition problems
are not applicable here.

Naively, one may interpret $\bar\kappa$ as the variance of the
end-to-end distance of the center-of-mass (CM) chain ${\bf X}(i)=[{\bf
  r}_1(i)+{\bf r}_2(i)]/2$.  Therefore, the $N$ dependence of
$\bar{\kappa}$ is given by the size
$R$ of the CM chain, with a scaling behaviour $R\sim N^{\nu}$.  
However, the CM chain is not expected to behave like an ordinary polymer,
 except in special situations, but in case it does so,
the size exponent $\nu=1/2$ for a Gaussian chain and $\nu\approx 0.588$
for a self-avoiding walk (polymers in good solvent).  For a
semiflexible Gaussian chain, the crossover from $R\sim N$ for $N\sim
l_p$ to $R\sim \sqrt{N}$ for $N\gg l_p$, is given by\cite{floryrev}
\begin{equation}
 \bar{\kappa}= 4R^2= 8l_p N\left\{ 1-
  \frac{l_p}{N}\left(1-e^{-N/l_p}\right)\right\}, 
  \label{eq:2}
\end{equation}
with $l_p$ as the persistence length.  Here both $l_p$ and length $N$
are measured in units of bond lengths which is set to one.  This
formula is used often for DNA when the CM chain more or less coincides
with the strands, i.e., in absence of bubbles.  In general, the
temperature dependence of ${\bar{\kappa}}/N^{2\nu}$ would show us how
DNA softens as the melting point $T_c$ is reached.

In terms of the individual coordinates, 
\begin{eqnarray}
\bar{\kappa}&=&2 \langle
r_1(N)^2\rangle_c \left(1+\frac{\langle  {\bf r}_1(N)\cdot{\bf
    r}_2(N)\rangle_c}{\langle  r_1(N)^2\rangle_c}\right ),
  \label{eq:3}
\end{eqnarray}
is determined by the inter-chain correlation.  In the perfect bound
state with no bubbles, ${\bf r_1 = r_2}$, and we get
$\bar{\kappa}/{\langle r_1(N)^2\rangle_c}=4$, while in the
high-temperature phase, if the two chains remain uncorrelated, then
$\bar{\kappa}$ is equal to the sum of the individual modulus.  Then,
$\bar{\kappa}/{\langle r_1(N)^2\rangle_c}=2$, for Gaussian chains.
The ratio is expected to be $>2$ for strands with excluded volume
interactions because there will be inter-strand correlations for long
chains as dictated by the second virial coefficient (or overlap
concentration $c^*$) \cite{degenne}.  Moreover, in the bound phase,
individual strands also acquire the stiffness of the state, punctuated
by bubbles and Y-fork.  Therefore, the microscopic stiffness would no
longer be the sole parameter determining the overall elasticity of the
chain (Fig.~\ref{fig:1}). In our study, it is assumed 
that the bending rigidity is isotropic i.e. the bending energy only depends 
on the angle by which the polymer is bent locally, and do not depend on the 
direction of bending, although, it has been shown that the bending rigidity 
in the direction of the grooves is essentially smaller than in the perpendicular 
direction\cite{olson,zhurkin}. To characterize the transition, we computed 
the fraction $f_b$ of broken bonds in the bubbles and $f_Y$, the fraction in 
the Y-fork like region. The transition temperature was determined from the 
specific heat curves; see Appendix A. In all cases, the order of transition 
is found to be independent of the value of stiffness $\eta$.

\section{Simulation Algorithm}

For simulation, we have used the zero parameter version of the
$\texttt{flatPERM}$ (Pruned and Enriched Rosenbluth Method) which
generates equilibrium configurations through cloning and
pruning\cite{prellberg,causo}. Both the strands are grown
simultaneously by considering all the joint possibilities of taking
steps together. The weighted atmosphere at each step, i.e., the number
of free sites available for the next step, serves as the weight of
that step $w_n$, and the weight of a configuration is the successive
multiplication of the weights of the previous steps $W_N=\prod_{n=1}^N
w_n$. For example, for the first step, each of the chains has $6$
different possibilities to step into.  Of a total of $36$
possibilities, there are $6$ possible ways of making a contact; thus
the local weighted atmosphere becomes $w_1=30+6\exp{(\epsilon/k_BT)}$.
Similarly, the weight for the second step including a bend and
excluded volume interaction is $w_2 = 4 \exp(\epsilon/k_BT) +
\exp(\epsilon/k_BT)\exp(\eta/k_BT) + 20$.  For Gaussian strands,
reverse steps in the ds mode with $E_b=\eta$ are considered with an
appropriate change in $w_2$.  The partition function for chain length
$n$ is estimated by averaging over the weights of configurations of
length $n$ with respect to the number of started tours where a tour is
a set of chains generated with a rooted tree topology between two
successive return to the \texttt{main()} function.  An average over
$10^{7}$ tours were used in this study for chain lengths
up to $2000$  and error bars are estimated on the fly; see 
Appendix D.  Pruning and enrichment is done continuously depending on
whether the ratio of the weight of the particular configuration $W_n$
and the partition function estimate $Z_n$ for length $n$,
$r=W_{n}/Z_{n}$ is smaller or greater than $1$ respectively. For ratio
$r<1$ the configuration keeps on growing with probability $r$ and
pruned otherwise.  While for $r>1$ we make $c$ distinct copies with
$c=min(\lfloor r \rfloor,a_n)$, where $a_n$ is the total atmosphere
$(a_n=a_1\times a_2)$ and each copy with weight $\frac{1}{c} W_n$. And
for $r=1$ the configuration continues to grow without any pruning or
enrichment. The input parameters for the simulations consist of the
temperature $T$, contact energy $\epsilon$, bending energy constant
$\eta$ with $\epsilon=k_B=1$ throughout the simulation unless
otherwise specified. To  translate  $\Delta T$ in our simulation 
to a variation in $\degree$C in  experiments, one  requires a proper scaling, 
e.g. $\epsilon$ can be estimated by comparing our $T_c$ to the experimental melting temperature.

 \begin{figure}[t]
\begin{center}
{\includegraphics{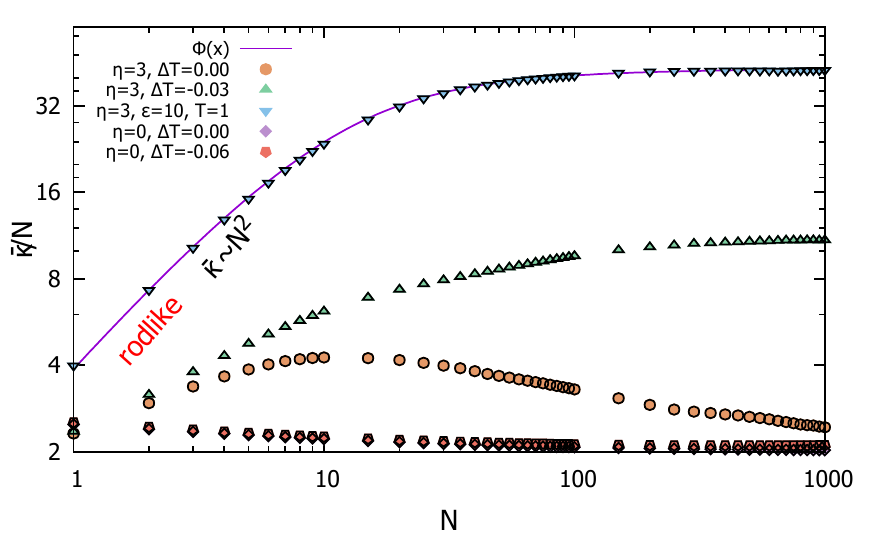}}
\hspace*{.5cm}(a) 
{\includegraphics{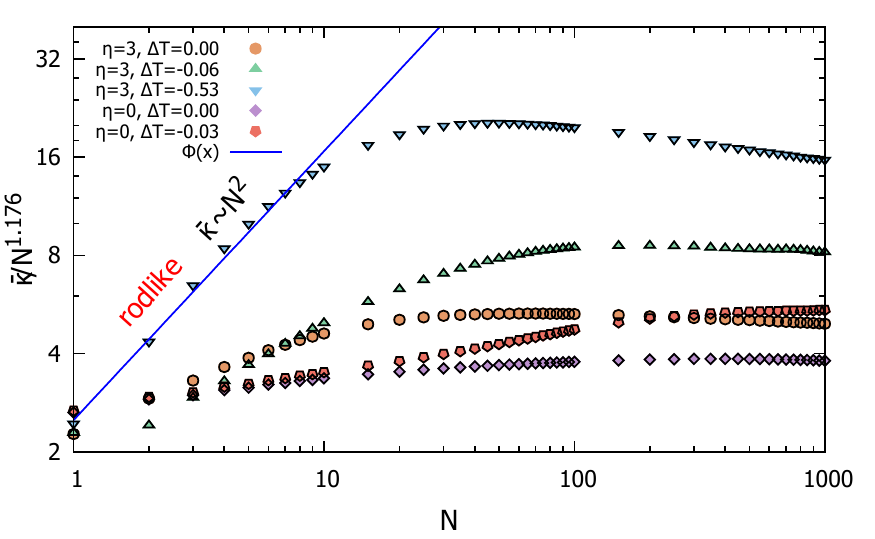}}
\hspace*{.5cm}(b)
\end{center}
\caption{(Color online) Log-log plot of $\bar\kappa/N^{2\nu}$ vs.
  $N$. (a) Model $\RN{1}$ for flexible $\eta=0$ and semi-flexible
  chain $\eta=3$, and different $\Delta T=T-T_c$. The curve $\phi(x)$ is
  a fit to the data points for $\epsilon/T=10, \eta/T=3$ using
  Eq.~\ref{eq:2} with $l_p$ as a parameter. (b) Model $\RN{2}$ for
  flexible $\eta=0$ and semi-flexible chain $\eta=3$, and different
  $\Delta T=T-T_c$. $\phi(x)$ is a straight line of slope ${(2-2\nu)}$
  representing the rodlike scaling regime. No other data sets show
  the initial slope of $\phi(x)$.
}
\label{KvsN}
\end{figure}

\section{Results and Discussion}
If dsDNA behaves as a semiflexible chain, then for small $N$, $\bar{\kappa}
\sim N^2$ as a rigid rod, with a crossover to Gaussian or SAW-like behaviour
for large $N$.
No rodlike behaviour is seen for $\eta=0$. Fig.~\ref{KvsN}(a) shows that for model 
\RN{1}, a tightly bound DNA (without bubbles) at $\epsilon/T=10, \eta/T=3$ ($k_B=1$) 
satisfies Eq.~(\ref{eq:2}) with $l_p=5.2$, consistent with the estimate of $l_p$
from a transfer matrix calculation (see Appendix F).  For model \RN{2} also, at low
temperatures, it is possible to define a rodlike behaviour; see  Fig.~\ref{KvsN}(b). 
However, the crossover description fails near the transition because of
substantial contributions from $f_b$ and/or ${f_Y}$.  In the log-log
plot, the slope for small lengths is not consistent with the rigid rod
expectations.  For $T$ close to $T_c$, DNA is neither rodlike nor
completely flexible for small chain lengths. We call this region as
soft DNA.  It follows that though an effective elastic constant can be
defined, persistence length from tangent correlations may not have any special
significance.

\begin{figure}[b]   
\begin{center}
  \includegraphics{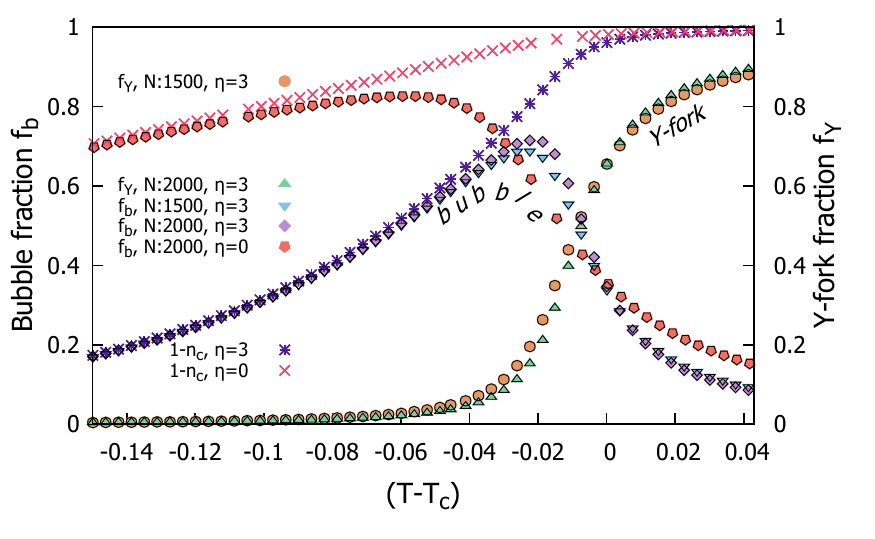}
  \hspace*{.5cm}(a)
  \includegraphics{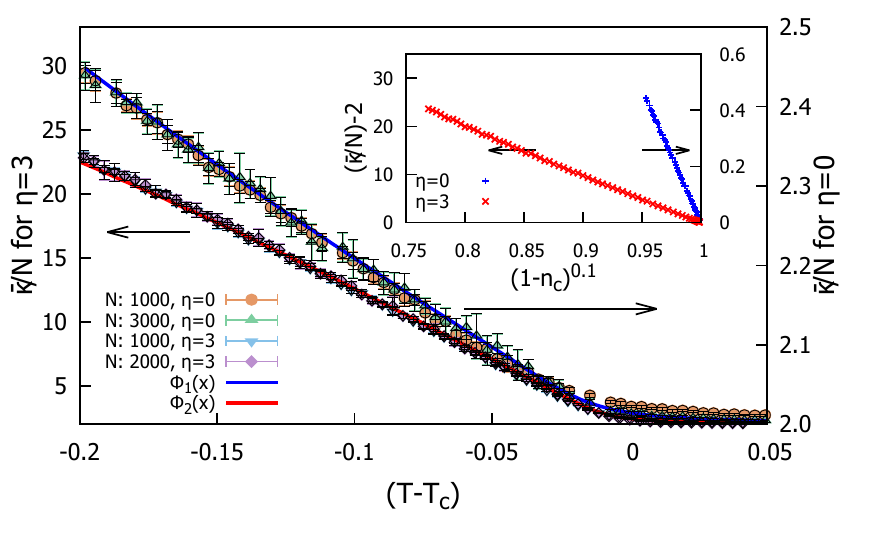}
  \hspace*{.5cm}(b)
\end{center}
\caption{(Color online) Bubble, Y and $\kappa$  for $\eta = 0, 3$ for Model $\RN{1}$ 
(continuous transition). (a) Bubble fraction $f_b$ , Y-fraction $f_Y$ , and total 
fraction $1 - n_c$ vs $(T - T_c )$. See text for the values of $T_c$. (b) Rescaled 
elastic modulus $(\bar{\kappa}/N^{2\nu} )$ vs $(T - T_c )$, with $\nu=1/2$. Eq. (3) 
is shown by solid lines, $\phi_1$ for $\eta = 0$ and $\phi_2$ for $\eta = 3$, with 
$n_c$ from (a). (Inset in (b)) Plot of $(\bar\kappa/N ) - 2$ vs $(1 - n_c )^{0.1}$ 
for $\eta = 0$ and $3$.}
\label{fby}
\end{figure}

Model $\RN{1}$: For Gaussian chains, the melting transition is
continuous at $T_c=1.336\pm0.006$ for $\eta=3$ and $T_c=0.928\pm0.006$
for $\eta=0$.  Below melting, bubbles develop, and the fraction of
broken bonds, $1-n_c$, increases with temperature continuously to 1 as
$T\to T_c-$ for $N\to\infty$.  For finite chains, there are also
broken bonds at the open end, but $f_Y$ vs $T$ curve sharpens into a
step function for $N\to\infty$.  Stiffness on the ds segments has the
effect of suppressing the bubble formation at lower temperatures but
the continuous transition remains intact.  Fig.~\ref{fby}(a) shows the
fractions for $\eta=0$ and $\eta=3$.  

These results are consistent with the Poland-Scheraga picture of DNA melting,
that most of the broken bonds are in the bubbles, while  the fraction in
the Y-region increases for $T>T_c$.    As the bubbles act like hinges, the 
decrease of the elastic constant with $T$ can be attributed to the broken 
bonds, thereby renormalizing the effective elastic constant as in Fig.~\ref{fig:1}.
As melting is continuous (see Appendix A and Fig.~\ref{fby}(a)) the change in the
elasticity near  melting is expected to be a power law
$\delta\kappa \equiv \bar{\kappa}/N - (\bar{\kappa}/N)_{\rm u} \sim n_c^q$ for 
$T\to T_c-$ \cite{comm2}. The order parameter has the asymptotic behaviour
$n_c\sim \left|(T-T_c)/T_c\right|^{\beta}$, so that the temperature dependence
of elastic constant is $\delta\kappa\sim \left|(T-T_c)/T_c\right|^{q\beta}$.
To extend the range of the asymptotic form valid for $T\to T_c$, we make an ansatz
\begin{equation}
  \label{eq:4}
\frac{1}{N}{\bar \kappa}= -\Delta_{\RN{1}}[(1-n_c^q)^{a}-1]+2, ~~~(\text{model}~\RN{1})
\end{equation}
where $\Delta_{\RN{1}}$ is the amplitude, and the exponents  $q$ and $a$  
take care of the softening by the bubbles. The values of  $q=1$ and $a=0.1$ 
are found to give a good agreement of the data shown in Fig.~\ref{fby}(b) when 
the values of $f_b,f_Y$ were used from Fig.~\ref{fby}(a).  The same picture 
remains valid for both $\eta=0$ and $\eta=3$.  We find  $\Delta_{\RN{1}}=9.5$ 
for $\eta=0$ and $\Delta_{\RN{1}}=98$ for $\eta=3$. This proposed model for 
elasticity suggests that the rigidity follows the order parameter curve as the 
melting point is approached from the bound side.

\begin{figure}[t]   
\begin{center}
  \includegraphics{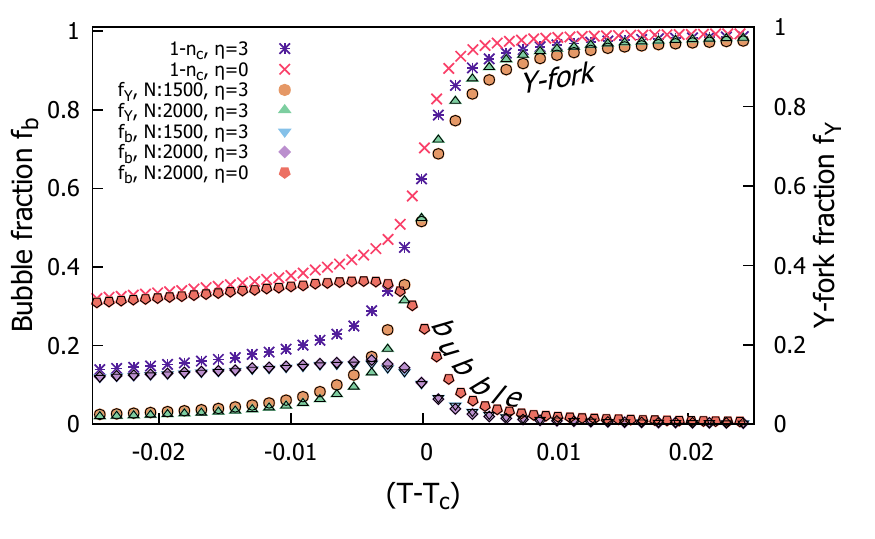}
  \hspace*{.5cm}(a)
  \includegraphics{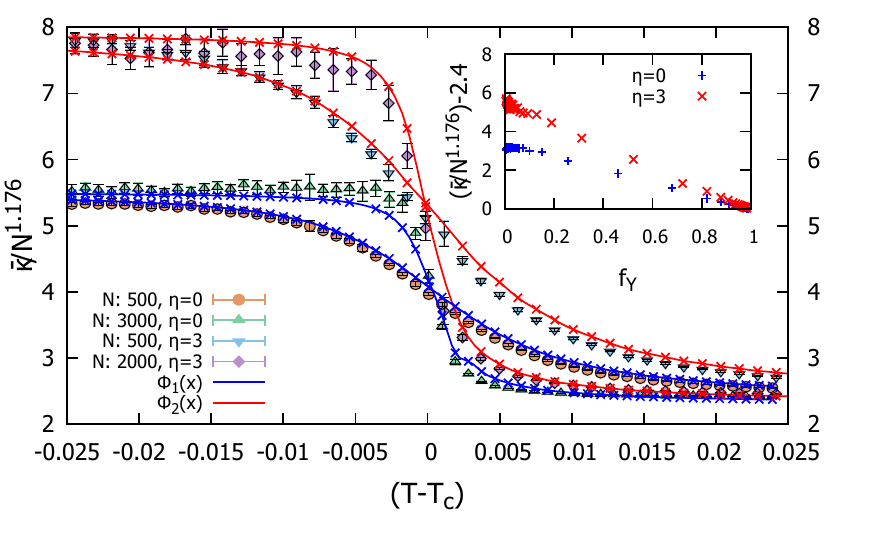}
  \hspace*{.5cm}(b)
\end{center}
\caption{(Color online) Same as in Fig.~\ref{fby} but for Model $\RN{2}$ with first 
order transition and $\nu=0.588$. (a) $f_b, f_Y, 1-n_c$ vs $(T-T_c)$. In (b) $\nu=0.588$ 
has been used, and Eq.~\ref{eq:5} is shown  by solid lines, $\phi_1$ for $\eta=0$ and 
$\phi_2$ for $\eta=3$, with $f_Y$ taken from (a). (Inset in (b) ) Plot of $(\bar\kappa/N^{1.176} ) - 2.4$ vs $f_Y$ for $\eta = 0$ and $3$.}
\label{fby2}
\end{figure}

Model \RN{2}: For self and mutually-avoiding chains the melting transition is 
first-order at $T_c=1.536\pm0.006$ for $\eta=3$ and $T_c=0.745\pm0.006$ for $\eta=0$. The 
temperature dependence of $f_b$ and $f_Y$ are shown in Fig.~\ref{fby2}(a), while that 
of $(\bar{\kappa}/N^{2\nu})$ in Fig.~\ref{fby2}(b), where $\nu$ takes into account 
the effect of excluded volume interaction (see Eq.~(\ref{eq:3})). There are significant 
differences from model \RN{1}.  Close to melting, most of the broken bonds are in the Y-fork, 
the fraction in the bubbles remains more or less the same. Consequently, Eq.~(\ref{eq:4}), 
encoding Fig.~\ref{fig:1}, is not meaningful; but instead an empirical equation, reminiscent 
of the lever rule in phase coexistence, is found to describe the data. We see  a deviation from 
the Poland-Scheraga picture. Our proposed model for rigidity in this case is based on the 
superposition of the elastic constant for the bound and the unbound Y-part in the proportion 
of $(1-f_Y):f_Y$ as
\begin{equation}
\label{eq:5}
\bar{\kappa}/N^{2\nu}=-\Delta_{\RN{2}}~{f_Y}^{2\nu}+(\bar{\kappa}/N^{2\nu})_{\rm b},~~~(\text{model}~\RN{2})
\end{equation}
which gives $(\bar{\kappa}/N^{2\nu})_{\rm b}$ of the bound phase for
${f_Y}\xrightarrow{} 0$, and $(\bar{\kappa}/N^{2\nu})_{\rm u}$ of the
unbound phase for ${f_Y}\xrightarrow{} 1$ with
$\Delta_{\RN{2}}=(\bar{\kappa}/N^{2\nu})_{\rm
  b}-(\bar{\kappa}/N^{2\nu})_{\rm u}$, the two limiting values were
adjusted to get a good fit with the values of $f_Y$ taken from Fig.~\ref{fby2}(a). 
These points  are also shown in the figure. The bubble contribution in the bound part 
is found to be very small. The  temperature independence of the elastic constant on the bound side 
away from melting is consistent with the experimental results of \cite{brunet}.  
The major implication of Eq.~(\ref{eq:5}) is the coexistence of the bound and the 
unbound state, although dsDNA is a single phase.

We note that the bubble fraction $f_b$ is lower for the semi-flexible
models [Figs.~\ref{fby}(a) and \ref{fby2}(a)], and  is a 
manifestation of the coupling between bubble
formation and DNA bending energetics (see Fig.~\ref{fig:2}). 
For stiffer bonds, it becomes energetically favourable to
maintain a bound state and make a straight move than to form a bubble to
become flexible; in other words, bending energy of a semi-flexible DNA
reduces the possibility of bubble formation.  This tendency to
maintain the bound state decreases the entropy of the system compared
to the $\eta=0$  case, thereby providing thermal stability to the
bound phase. Thus the melting temperature is higher for nonzero $\eta$,
and the transition becomes sharper.  However, the bubble size
distribution and thus the average bubble length 
near the transition remain unaltered by stiffness.

\section{Conclusion}
We conclude by comparing the melting picture of DNA with that of a
crystal.  Two main contenders of the mechanism for melting of a
crystal are the homogeneous melting via the formation of defects,
topological or nontopological, and a surface melting\cite{yip,mei}. If
the defects form anywhere in the bulk crystal due to thermal
fluctuations, the ordering is destroyed with a reduction of the
rigidity.  The melting process is then homogeneous.  A different
possibility is a surface melting where a wetting liquid layer is
formed on the surface and the thickness of the layer diverges at the
melting point.  We do see analogs of these two processes in DNA
melting, 
though distinctly different in detail, and 
dependent on the order of the transition.  For continuous transitions,
it is the Poland-Scheraga scheme of homogeneous bubble formation that
modifies the elastic constant as in Eq. (\ref{eq:4}).  For a ribbon
picture to be applicable, a coarse-graining as shown in
Fig.~\ref{fig:1}, is necessary. The reduction in rigidity follows the temperature
 dependence of the fraction of intact bonds. On the other
hand, with excluded volume interaction the melting process starts at
the open end, like surface melting, at temperatures below the real
melting temperature, forming the Y-region.  The melting process
completes when the length of the Y-region diverges (for infinite
chains). In this scenario, for long chains, the density of broken
bonds when expressed in terms of the length of the bound segment, viz.,
$f_b/(1-f_Y)$ should be independent of $N$, as we found to be the case
(see Appendix E).
This picture also suggests that if a dsDNA is capped by a sequence
of high binding energy (i.e.  of a higher melting point), then there is
a possibility of superheating a dsDNA, when the dsDNA state can be
maintained in its bound phase above the melting point.  This
nonequilibrium aspect is beyond the scope of  equilibrium
simulations.

\section*{\label{acknow}Acknowledgements} 
Computations performed on a FUJITSU  high-performance computing facility   ({\it SAMKHYA}) 
at the Institute Of Physics Bhubaneswar. SMB acknowledges support from the  JCBose Fellowship grant.

\appendix

\begin{figure}[b]
    \centering
    \includegraphics[width=0.47\textwidth]{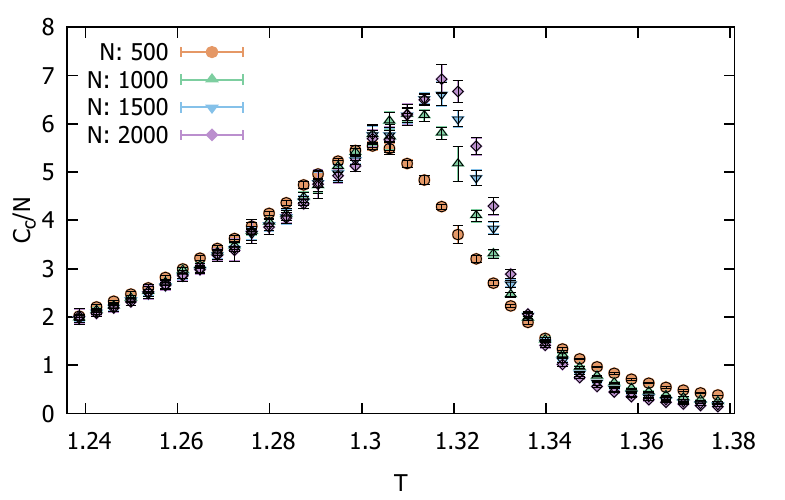}
\hspace*{.5cm}(a)    
    \includegraphics[width=0.47\textwidth]{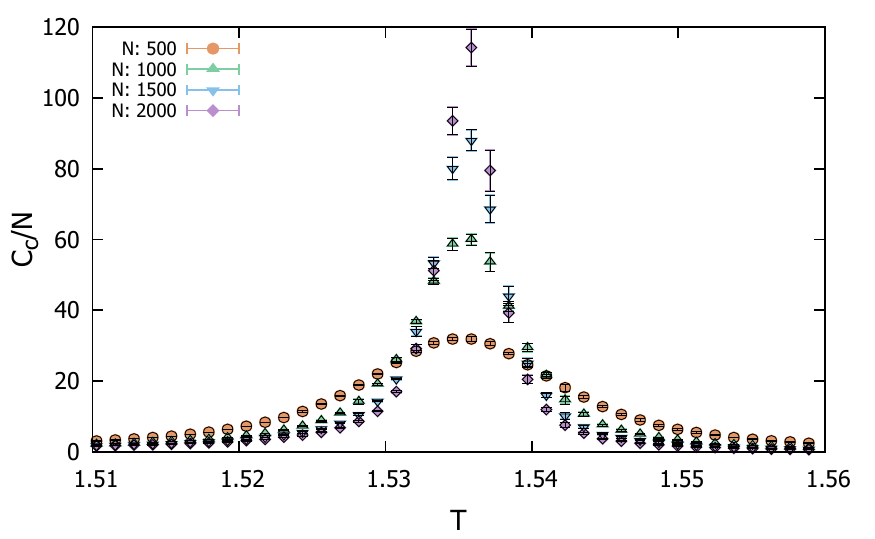}
\hspace*{.5cm}(b)
    \caption{(Color online) Contact number fluctuation. (a) Model
      $\RN{1}$. (b) Model $\RN{2}$. Both results are for semi-flexible
      models with bending energy constant $\eta=3$.}
    \label{fig:my_label1}
\end{figure}{}

\section{Estimation  of the Transition point}
The transition (melting) point has been estimated from the contact
number fluctuation per basepair $C_c/N$ (related to the specific
heat).
For model $\RN{1}$, undergoing a continuous
transition, the transition point is determined from  the intersection
point of the curves for 
various lengths $N$, that remains invariant under a change of system
size (Fig.~\ref{fig:my_label1}(a))
and the transition point is estimated
to be $T_c=1.336\pm 0.006$ for $\eta=3$ .    For model $\RN{2}$ which undergoes
a first-order transition,  the transition point is determined from the
peak of the contact number fluctuation curves
(Fig.~\ref{fig:my_label1}(b)), and the transition point 
is estimated to be $T_c=1.536\pm 0.006$ for $\eta=3$.

\section{Bubble Size Distribution}
\begin{figure}[h]
   \includegraphics[width=0.5\textwidth]{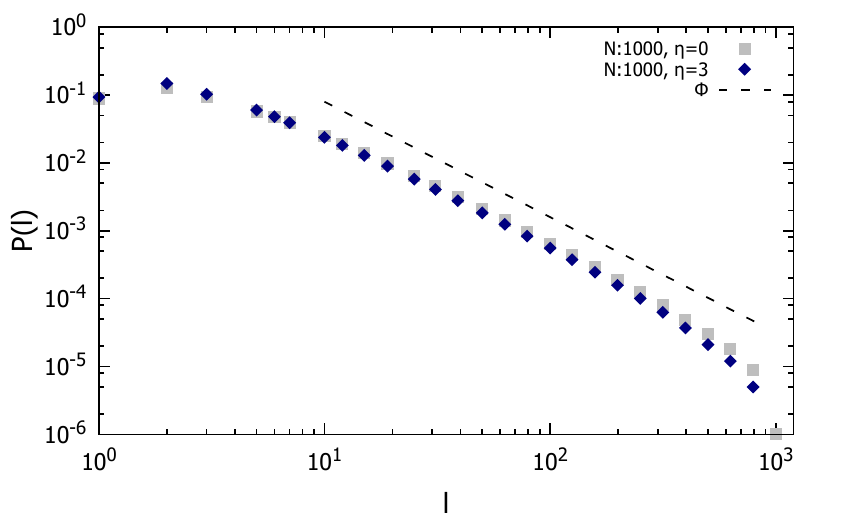}
\hspace*{.5cm}(a)    
   \includegraphics[width=0.5\textwidth]{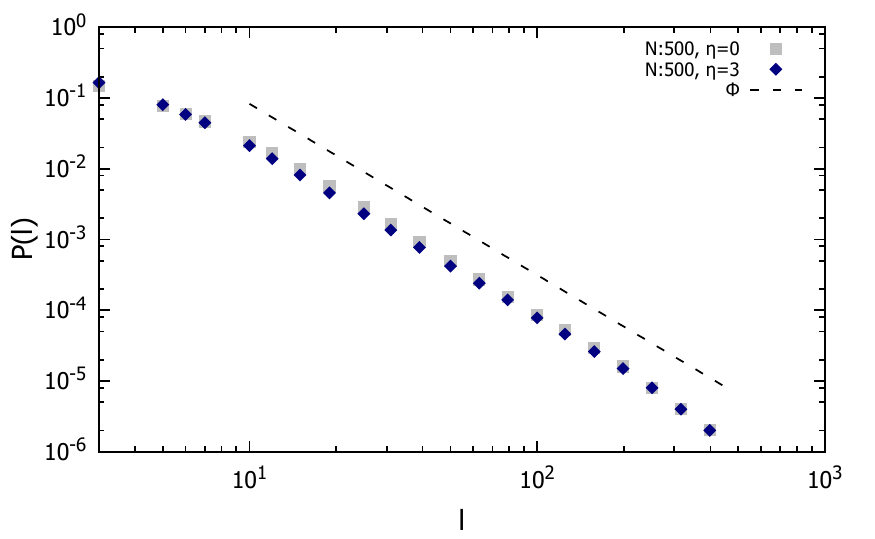}
\hspace*{.5cm}(b) 
   \caption{(Color online) Bubble size distribution at the melting point.  (a) Model
     $\RN{1}$ for $\eta=0, 3$ and $T \approx T_c=.928, 1.33$
     respectively, for a chain length $N=1000$.  The straight line
     $\phi(x)\sim x^{-1.7}$ is a fit to the intermediate region of the $\eta=0$
     data points. (b) Model $\RN{2}$ for $\eta=0, 3$ at $T \approx
     T_c=.75, 1.54$ respectively, for chain length $N=500$. The
     straight line $\phi(x)\sim x^{-2.4}$ is a fit to the intermediate
     region of the $\eta=3$ data points.}
    \label{fig:my_label2}
\end{figure}

The bubble size distribution at the transition point scales as\cite{fisher,carlon}
\begin{equation}
    P(l)\sim l^{-\psi}
\end{equation}
where $P(l)$ is the probability of a bubble of length $l$  and the
exponent $\psi$ is related 
to the nature of the transition\cite{fisher}.
If $\psi>2$, it represents a first-order transition,
while $1<\psi<2$ represents a continuous transition and for $\psi<1$ there
is no transition at all. As per the definition of a bubble, broken
bonds in the Y-fork is not included in the bubble statistics. See
Fig.~\ref{fig:my_label2}.  The observed slopes are consistent with a
continuous transition for Model $\RN{1}$ (Gaussian chains) but a
first-order transition for Model $\RN{2}$ (chains with excluded volume
interaction).
  
{}

\section{Benchmark for Simulation results}
\begin{figure}[t]
    \centering
    \includegraphics[width=0.5\textwidth]{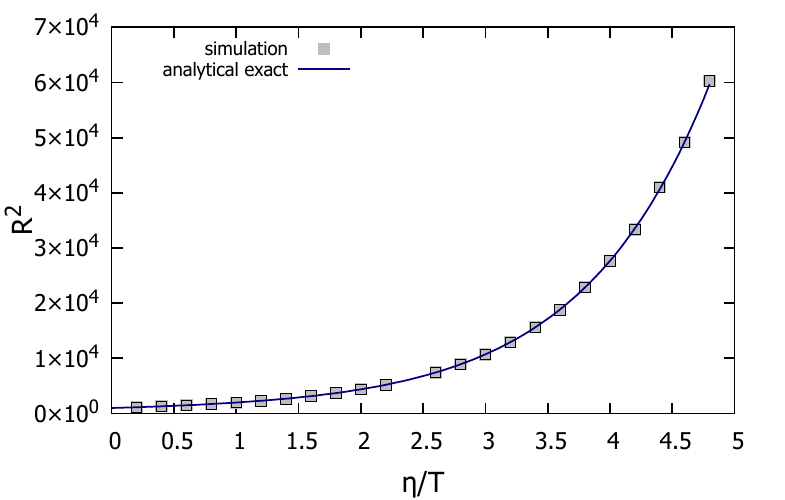}
    \caption{(Color online) The mean squared  end-to-end distance $R^{2}$ vs
      $\bar\eta$, obtained both analytically (Eq. (\ref{eq:s1})), and
      from simulation at temperature $T$ much below the $\eta=0$
      transition point $T_c$, $\epsilon/T=10 \gg \epsilon/T_{c}=
      1.077.$  This is  for length $N=1000$. } 
    \label{fig:my_label3}
\end{figure}

To check for the accuracy of our code in case of semi-flexible
Gaussian chains, we calculated the mean squared end-to-end distance
for different rigidities at a very low temperature where the DNA is in
the bound state, and compared with the exact analytical result for a
single ideal semi-flexible chain. Analytically, for an ideal
semi-flexible chain the mean squared end-to-end distance varies with
the rigidity of the local bending as\cite{floryrev}
 \begin{equation}
\label{eq:s1}
\langle R^{2}\rangle = b^{2}(N+1)\frac{1+\textit{L}({\bar\eta})}{1-\textit{L}({\bar\eta})} - 2b^{2}\textit{L}({\bar\eta})\frac{1-\textit{L}^{N+1}({\bar\eta})}{(1-\textit{L}({\bar\eta}))^{2}},
\end{equation}
where
$$\textit{L}({\bar\eta})=\frac{\partial}{\partial \bar{\eta}}\ln Z(\bar{\eta})= \frac{\sinh({\bar\eta})}{\cosh({\bar\eta})+2},$$
${\bar\eta}=\eta/k_{B}T$ and $Z(\bar{\eta})$ is the two step partition function 
when no external force is applied.  The comparison is shown in Fig.~\ref{fig:my_label3}. 
The form of $L({\bar\eta})$ is specific to our convention of the
energetics for polymer bending, where if overlapped (i.e. in the
single chain limit) the polymer has the following local partition
function for a two step walk
\begin{equation}
Z=e^{{\bar\eta}}+e^{-{\bar\eta}}+4    
\end{equation}
where $e^{{\bar\eta}}$ is for moving straight, $e^{-{\bar\eta}}$ for
moving backwards and 1 for right angle turns over a cubic lattice
according to the bending energy $E_b=-\eta \cos{\theta}$.

\begin{figure}[t]
    \centering
    \includegraphics[width=0.5\textwidth]{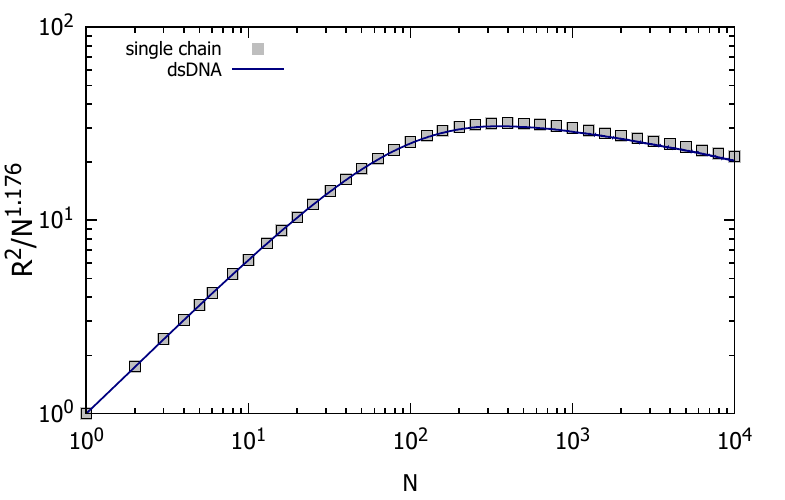}
    \caption{(Color online) Log-log plot of the scaled mean squared end-to-end
      distance $R^{2}/N^{2\nu}$ vs. the chain length $N$ for
      model $\RN{2}$, with Boltzmann factors
      $\exp{(\epsilon/T)}=7\times 10^4$ for contact and
      $\exp{(-\bar\eta)}=0.005$ for bends, for the dsDNA, and a single
      self-avoiding semi-flexible chain with the same Boltzmann factor
      for bending $\exp{(-\bar\eta)}=0.005$.  }  \label{fig:my_label4}
\end{figure}

In the presence of self and mutual avoidance, for model $\RN{2}$ of
DNA, a good check would be to compare the size of the polymer chain at
a very low temperature (i.e.  large $\epsilon/T$) when the two chains
would be completely in the overlapped state and would behave as a
single rigid chain, with that of the single chain of the same
rigidity\cite{hsuS}.  The comparison is shown in Fig.~\ref{fig:my_label4}. 
Note that for both dsDNA and a single
self-avoiding chain, the relatve weight for bending is taken as 
$\exp(-\bar\eta)=0.005$.  

\section{On the fly error calculation for fluctuating quantities}
The estimation of error for any thermodynamical observable
is the fluctuation of that quantity. For quantities
which are fluctuating in itself e.g. contact number
fluctuation $C_c$ or elastic modulus $\kappa$ estimation of error becomes tricky. The
way PERM is implemented every tour provides an independent
estimate of any quantity which contributes to
another sample in the running average. Now, with every
new estimate from a tour the difference of the present estimate
from the estimate of the average up to now gives
a measure for the fluctuation of that quantity. The updates
of the mean and the fluctuation of a quantity $x$
follows the scheme
\begin{eqnarray}
     \overline{x}_n&=&\overline{x}_{n-1}+(x_n-\overline{x}_{n-1})/n\\
     d_n^2&=&d_{n-1}^2+(x_n-\overline{x}_n)(x_n-\overline{x}_{n-1}),
\end{eqnarray}
where $x_n$ is the current $n$th estimate of the quantity,
$\overline{x}_{n-1}$ represents the average up to $n-1$ samples and
$d_n^2=\sum_{i=1}^n\left(x_i-\overline{x}_n\right)^2$. Thus, the
standard deviation is given by
\begin{equation}
S_n=\sqrt{\frac{d_n^2}{(n-1)}}.
\end{equation}
This method is known as Welford's method. 

\begin{figure}
    \centering
    \includegraphics[width=0.5\textwidth]{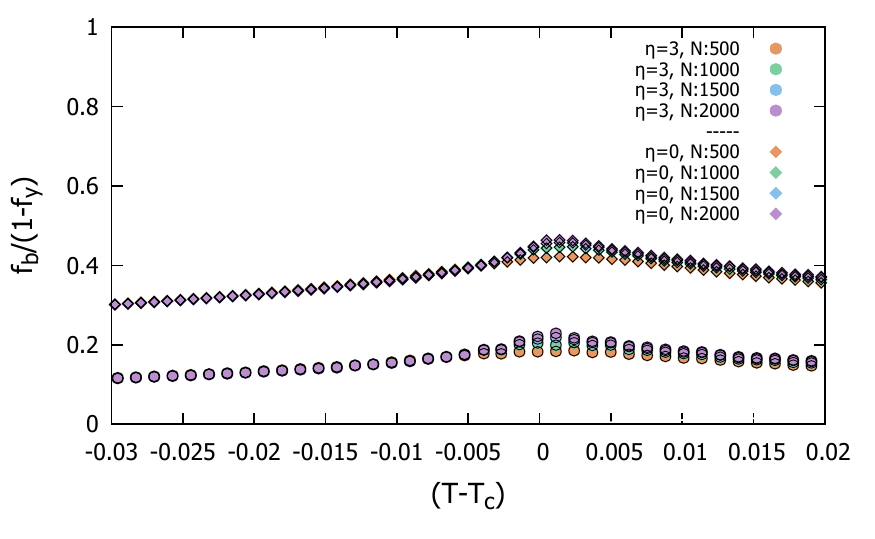}
    \caption{(Color online) Plot of $n_b=f_b/(1-f_Y)$ vs $T-T_c$ for $\eta=0, 3$ and
      various $N$, using the data of Fig.~\ref{fby2}(a). The data
      collapse corroborates the picture of phase coexistence.  The
      deviations near $T=T_c$ is the usual finite size effect at a
      phase transition.  }  \label{fig:my_label5}
\end{figure}

\section{Data collapse for first-order melting }
Let $N_b$ and $N_Y$ be the total number of broken bonds in bubbles and
the Y-fork region. Then $f_b=N_b/N$ and $f_Y=N_Y/N$ and their
temperature dependence is shown in Fig.~\ref{fby2}(a).  If we treat the
chain as consisting of two segments, bound and Y, then the density of
broken bonds in the bound segment will be
$n_b=N_b/(N-N_Y)=f_b/(1-f_Y)$. For long bound segments, $n_b$ should
be independent of $N$.  The plot of $n_b$ vs $T-T_c$ in Fig.~\ref{fig:my_label5} shows a nice collapse both for $\eta=0$ and
$\eta=3$, except for the usual finite size effect near $T_c$.  This
data collapse validates the coexistence picture.

\section{Transfer Matrix Calculation of Persistence Length }
For Gaussian semiflexible chains, the tangent-tangent correlation or
bond-bond correlation (Fig.~\ref{fig:1}) decays exponentially for
large $|i-j|$, $\langle{\bf t}_i\cdot{\bf t}_j\rangle\sim
\exp(-|i-j|b/l_p)$ providing a definition for persistence length $l_p$, where $b$ is the bond length.
This definition is not applicable for cases with excluded volume
interaction, as SAWs are critical objects \cite{schafer}.  
  The persistence length for
model \RN{1} with $\eta$ at temperature $T$ ($\bar\eta=\eta/k_BT$) can
be exactly calculated from transfer 
matrix calculation of a two step walk. The transfer matrix for a two
step walk is written as
\begin{equation}
  \begin{pmatrix}
e^{\bar\eta} & 1 & 1 & 1 & 1 & e^{-{\bar\eta}}\\
1 & e^{\bar\eta} & 1 & 1 & e^{-{\bar\eta}} & 1\\
1 & 1 & e^{\bar\eta} & e^{-{\bar\eta}}& 1 & 1\\
1 & 1 &  e^{-{\bar\eta}} & e^{\bar\eta} & 1 & 1\\
1 &  e^{-{\bar\eta}} &1& 1 & e^{\bar\eta} & 1\\
e^{-{\bar\eta}} & 1 & 1 & 1 & 1 & e^{\bar\eta}

\end{pmatrix}  
\end{equation}
The largest eigenvalue for the above matrix is $\lambda_1=4+2
\cosh{\bar\eta}$, and for $\bar\eta=3$ the second largest eigenvalue obtained
using \texttt{MATHEMATICA} is $\lambda_2=19.562$.
Therefore, the persistence length is $
l_p={[\ln(\lambda_1/\lambda_2)]}^{-1}\approx 5.3$.


\vspace{1cm}


\begin{thebibliography}{99}

\bibitem{purohit} P.K. Purohit, J. Kondev, and R. Phillips. Mechanics of  DNA packaging in viruses.  {\it Proc. Natl. Acad. Sci. USA} {\bf 100}, 3173-3178 (2003).

\bibitem{garcia} H.G. Garcia, P. Grayson, L. Han, M. Inamdar, J. Kondev, P.C. Nelson, R. Phillips, J. Widom, and P.A. Wiggins. Biological Consequences of Tightly Bent DNA: The Other Life of a Macromolecular Celebrity. {\it Biopolymers} {\bf 85}, 115 (2007).


\bibitem{hori} M. Horikoshi, C. Bertuccioli, R. Takada, J. Wang, T. Yamamoto, and R.G. Roeder. Transcription factor TFIID induces DNA bending upon binding to the TATA element. {\it Proc. Natl. Acad. Sci. USA} {\bf 89}, 1060-1064 (1992).

\bibitem{rees} W. Rees, R. Keller, J. Vesenka, G. Yang, and C. Bustamante. Evidence of DNA bending in transcription complexes imaged by scanning force microscopy. {\it Science} {\bf 260}, 1646-1649 (1987).

\bibitem{topoiso} S. Lee, S-R. Jung, K. Heo, Jo Ann W. Byl, J.E. Deweese, N. Osheroff, and S. Hohng. DNA cleavage and opening reactions of human topoisomerase II$\alpha$ are regulated via Mg$^{2+}$-mediated dynamic bending of gate-DNA. {\it Proc. Natl. Acad. Sci. USA} {\bf 109}, 2925-2930 (2012).


\bibitem{mullis} K.B. Mullis and F.A. Faloona. Specific Synthesis of DNA in Vitro via a Polymerase-Catalyzed Chain Reaction. {\it Methods Enzymol} {\bf 155}, 335-350 (1987).

\bibitem{vologo} S. Geggier and A. Vologodskii. Sequence dependence of DNA bending rigidity. {\it Proc. Natl. Acad. Sci. USA} {\bf 107}, 15421-15426 (2010).


\bibitem{hager} P.J. Hagerman. Flexibility of DNA, {\it Ann. Rev. Biophys. Biophys. Chem.} {\bf 17}, 265 (1988).

\bibitem{dietler} K. Rechendorff, G. Witz, J. Adamcik, and G. Dietler. Persistence length and scaling properties of single-stranded DNA adsorbed on modified graphite. {\it J. Chem. Phys.} {\bf 131}, 095103 (2009).

\bibitem{abels} J.A. Abels, F. Moreno-Herrero, T. van der Heijden, C. Dekker, and N.H. Dekker.  Single-Molecule Measurements of the Persistence Length of Double-Stranded RNA. {\it Biophys. J.} {\bf 88},  2737 (2005).

\bibitem{brunet} A. Brunet, A. {\it et al.}. How does temperature impact the conformation of single DNA molecules below melting temperature? {\it Nucl. Acd. Res.} {\bf 46}, 2074 (2018).

\bibitem{olson} W.K. Olson, A.A. Gorin, X.J. Lu, L.M. Hock, and V.B. Zhurkin.  DNA sequence-dependent deformability deduced from protein-DNA crystal complexes. {\it Proc. Natl. Acad. Sci. USA} {\bf 95}, 11163-11168 (1998).

  
\bibitem{kivelson} S. Kivelson and S.A. Kivelson. Defining emergence in physics. {\it npj Quantum Materials} {\bf 1}, 16024 (2016).

\bibitem{schafer} L. Sch\"afer, A. Ostendorf and J. Hager. Scaling of the correlations among segment directions of a self-repelling polymer chain. {\it J. Phys. A: Math. Gen.} {\bf 32}, 7875 (1999).

\bibitem{hsu} H-P. Hsu, W. Paul and K. Binder. Standard Definitions of Persistence Length Do Not Describe the Local ``Intrinsic" Stiffness of Real Polymer Chains. {\it Macromolecules} {\bf 43}, 3094 (2010).

\bibitem{vologodski3} A. Vologodskii. {\it Biophysics of DNA} (Cambridge University Press, Cambridge, UK, 2015).


\bibitem{comm3} Our models differ from other models considered previously, e.g., in \cite{palmeri}, which considered a distance dependent elastic constant. Our models differ with \cite{palmeri} in the following respect: (i) Both models $\RN{1}$ and $\RN{2}$ incorporate  polymeric correlations in bubbles, that change the nature of the melting transition, but this correlation is not present  in the model of \cite{palmeri}.  (ii) Our  Model $\RN{2}$ considers excluded volume interaction, both inter and intra strand, which are known to be relevant interactions for polymers, and change the reunion exponents relevant for melting  \cite{fisher}.  This interaction is absent in \cite{palmeri}. 

\bibitem{palmeri} J. Palmeri, M. Manghi, and N. Destainville. Thermal Denaturation of Fluctuating DNA Driven by Bending Entropy. {\it Phys. Rev. Lett.} {\bf 99}, 088103 (2007).


\bibitem{fisher} M.E. Fisher. Walks, walls, wetting, and melting. {\it J. Stat. Phys.} {\bf 34}, 667 (1984).

\bibitem{causo} M.S. Causo, B. Coluzzi, and P. Grassberger. Simple model for the DNA denaturation transition. {\it Phys. Rev. E} {\bf 62}, 3958 (1999).

\bibitem{yan} J. Yan and J.F. Marko. Localized Single-Stranded Bubble Mechanism for Cyclization of Short Double Helix DNA. {\it Phys. Rev. Lett.} {\bf 93}, 108108 (2004).

\bibitem{forties} R.A. Forties, R. Bundschuh and M.G. Poirier. The flexibility of locally melted DNA. {\it Nucleic Acids Res.} {\bf 37}, (2009).

\bibitem{wiggins} P.A. Wiggins, R. Phillips and P.C. Nelson. Exact theory of kinkable elastic polymers. {\it Phys. Rev. E} {\bf 71}, 021909 (2005).

\bibitem{vologodskii2} A. Vologodskii, and M.D. Frank-Kamenetskii. Strong bending of the DNA double helix. {\it Nucleic Acids Res.} {\bf 41}, 6785 (2013).

\bibitem{bonnet} G. Altan-Bonnet, A. Libchaber and O. Krichevsky. Bubble Dynamics in Double-Stranded DNA. {\it Phys. Rev. Lett.} {\bf 90}, 138101 (2003).

\bibitem{palsmb} T. Pal, T. and S.M. Bhattacharjee. Rigidity of melting DNA. {\it Phys. Rev. E.} {\bf 93}, 052102 (2016).


\bibitem{chongli} C. Yuan, E. Rhoades, X.W. Lou and L.A. Archer. Spontaneous sharp bending of DNA: role of melting bubbles. {\it Nucleic Acids Res.} {\bf 34}, 4554 (2006).

\bibitem{ramstein} J. Ramstein and R. Lavery. Energetic coupling between DNA bending and base pair opening. {\it Proc. Natl. Acad. Sci. USA} {\bf 85}, 7231 (1988).

\bibitem{floryrev} S.M. Bhattacharjee, A. Giacometti and A. Maritan. Flory Theory for Polymers. {\it J. Phys. Condens. Matter} {\bf 25}, 503101 (2013).
 
\bibitem{comm2} Here $(\bar\kappa/N)_u$ refers to the unbound phase for $N\rightarrow \infty$.  
 
\bibitem{yip} S.R. Phillpot, S. Yip and D. Wolf. How Do Crystals Melt?. {\it Comp. in Phys.} {\bf 3}, 20 (1989).

\bibitem{mei} Q.S. Mei and K. Lu. Melting and superheating of crystalline solids: From bulk to nanocrystals. {\it Prog in Mat. Sci.} , {\bf 52} 1175-1262 (2007).

\bibitem{zhurkin} V.B. Zhurkin, Y.P. Lysov and V.I. Ivanov. Anisotropic flexibility of DNA and the nucleosomal structure. {\it Nucleic Acids Res.} {\bf 6}, 1081 (1979).

\bibitem{degenne} P.G. de Gennes, {\it Scaling Concepts in Polymer Physics} (Cornell University Press, Ithaca, NY, 1979).

\bibitem{prellberg} T. Prellberg and J. Krawczyk. Flat Histogram Version of the Pruned and Enriched Rosenbluth Method. {\it Phys. Rev. Lett.} {\bf 92}, 120602 (2004).

\bibitem{carlon} E. Carlon, E. Orlandini and A.L. Stella. Roles of stiffness and excluded volume in DNA denaturation. {\it Phys. Rev. Lett.}, {\bf 88}, 198101 (2002).

\bibitem{hsuS} H-P. Hsu, W. Paul and K. Binder. Polymer chain stiffness vs. excluded volume: A Monte Carlo study of the crossover towards the worm-like chain model. {\it Euro Phys. Lett.} {\bf 92}, 28003 (2010).


\end{thebibliography}
\end{document}